\documentclass[twocolumn,showpacs,preprintnumbers,amsmath,amssymb]{revtex4}

\usepackage{graphicx}
\usepackage{dcolumn}
\usepackage{bm}

\begin{document}

\title{Perturbation: the Catastrophe Causer in Scale-Free Networks}
\author{Tao Zhou}
\author{Bing-Hong Wang}
\email{bhwang@ustc.edu.cn, Fax:+86-551-3603574.}
\affiliation{%
Nonlinear Science Center and Department of Modern Physics,
University of Science and Technology of China, Heifei, 230026, PR
China
}%

\date{\today}

\begin{abstract}
A new model about cascading occurrences caused by perturbation is
established to search after the mechanism because of which
catastrophes in networks occur. We investigate the avalanche
dynamics of our model on 2-dimension Euclidean lattices and
scale-free networks and find out the avalanche dynamic behaviors
is very sensitive to the topological structure of networks. The
experiments show that the catastrophes occur much more frequently
in scale-free networks than in Euclidean lattices and the greatest
catastrophe in scale-free networks is much more serious than that
in Euclidean lattices. Further more, we have studied how to reduce
the catastrophes' degree, and have schemed out an effective
strategy, called targeted safeguard-strategy for scale-free
networks.
\end{abstract}


\maketitle

Many social, biological, and communication systems can be properly
described as complex networks with vertices representing
individuals or organizations and edges mimicking the interactions
among them. Recently, the ubiquity of a power-law degree
distribution in real-life networks has attracted a lot of
attention\cite{Reviews}. Examples of such networks (scale-free
networks or SF networks for short) are numerous: these include the
Internet, the World Wide Web, social networks of acquaintance or
other relations between individuals, metabolic networks, integer
networks, food webs, etc.\cite{Networks}. The ultimate goal of the
study of the topological structure of networks is to understand
and explain the workings of systems built upon those networks, for
instance, to understand how the topology of the World Wide Web
affects Web surfing and search engines, how the structure of
social networks affects the spread of diseases, information,
rumors or other things, how the structure of a food web affects
population dynamics, and so on.

The catastrophes in real-life networks can be see everywhere, such
as the traffic jams taking place in road-networks, the
communication congestions taking place in internet, the economic
crisis taking place in the network of financial institutions, and
so on. Therefore, it is not only of major theoretic interest, but
also of great practical significance to understand the mechanism
because of which those catastrophes occur. Intuitively, one may
consider the breakdown of plentiful vertices or edges at the same
time in networks to be the reason\cite{Albert 00}. He is
undoubtedly right as it is easy to imagine the damage of roads or
the failure of servers lead to a serious traffic jam or
communication congestion, respectively. However, in most
situations, the catastrophes surrounding us are not like that. In
the present letter, a man named ``{\bf Perturbation}" is caught,
who is indicted to be the causer in most catastrophes.

{\it Bianconi} and {\it Marsili} referred to an example about the
catastrophe caused by perturbation\cite{Bianconi 03}. The example
they mentioned is routing tables in the internet, which can be
considered as a dynamic communication network. In the beginning, a
change (perturbation) in some router's table may inadvertently
cause congestion at some node downstream. This may trigger several
other changes in that local neighborhood, as routers try to avoid
the congested node. But these changes may, in their turn, cause
further congestion elsewhere, and the problem may expand even
further, as a large avalanche (catastrophe), to a wider region.
Similar phenomena may take place in various networks.

{\it Bak}, {\it Tang} and {\it Wiesenfeld} introduced a so-called
sandpile model (BTW model) to explain such cascading occurrences
on networks, which is considered as a prototypical theoretical
model exhibiting the catastrophes (avalanche behavior) caused by
perturbation\cite{BTW}. However, BTW model is based on the
Euclidean lattices, which are very different from the reality for
the real-life networks that have power-law degree distribution. In
addition, the open boundary conditions make it hard to directly
extend BTW model onto SF networks. {\it Olami}, {\it Feder} and
{\it Christensen} established a model (OFC model) of earthquakes
in nonconservative systems\cite{OFC}, which may be more
appropriate to mimic the catastrophes in SF networks than BTW
model for all the real-life systems are nonconservative.

Recently, a few interesting and significant works about how the
topological structure of networks affects self-organized
criticality (SOC) based on BTW model or OFC model have been
achieved. {\it Lise, et al} investigated the OFC model on annealed
and quenched random networks\cite{Lise 9602}, {\it Arcangelis} and
{\it Herrmann} studied the BTW model on small-world
networks\cite{Arcangelis 01}. {\it Motter} and {\it Lai} studied
the cascade-based attacks on SF networks and have found that a
large-scale cascade nay be triggered by removing a single key
vertex\cite{Motter 02}. {\it Goh, et al} investigated the
avalanche dynamics on SF networks using BTW model and obtained the
exponent $\tau$ for the power-law avalanche size distribution and
the dynamic exponent $z$\cite{God 0304}. Their work concentrated
on the existence of SOC, thus they did not discuss whether the
catastrophes occur more frequently in SF networks than Euclidean
lattices. In addition, since the threshold value of each vertices
is assigned to be equal to its degree, one can not make sure which
(the power-law degree distribution, the power-law threshold height
distribution or both) is the main reason that lead to the
power-law distribution of avalanche size.

In the present letter, a model similar to OFC's is established to
mimic the catastrophes occurring in networks. We have found that
the catastrophes occur much more frequently in SF networks than in
Euclidean lattices and the greatest catastrophe in SF networks is
much more serious than that in Euclidean lattices. Further more,
we have studied how to reduce the catastrophes' degree, and have
schemed out an effective safeguard-strategy for SF networks.

In our model, to each vertices of the network is associated a real
variable $F_x$, which initially takes the value 0 and can be
considered as energy, tension, flux or some other things. At each
time step, a perturbation $\delta$ is added to a randomly chosen
vertex $x$, which means the variable $F_x$ increases by $\delta$,
where $\delta$ is randomly selected in the interval $(0,1)$. If
$F_x$ reaches or exceeds the threshold value $Z_x$, then the
vertex $x$ becomes unstable and the $(1-\varepsilon )Z_x$ energies
topple to its neighbor-nodes, with a small fraction $\varepsilon$
of energies being lost: $F_x\rightarrow F_x-Z_x$, and
$F_y\rightarrow F_y+(1-\varepsilon )Z_x/d(x)$ for all vertices $y$
adjacent to $x$, where $d(x)$ is the degree of vertex $x$ that
denotes the number of neighbor-vertices of $x$. The parameter
$\varepsilon$ controls the level of conservation of the dynamics
and it takes values between 0 and 1, where $\varepsilon =0$
corresponds to the conservative case. Here, to avoid the system
being overloaded in the end, $\varepsilon$ is always set to be
larger than 0. If this toppling causes any of the adjacent
vertices receiving energies to be unstable, subsequent toppling
follow on those nodes in parallel until there is no unstable node
left. This process defines an avalanche.

\begin{figure}
\scalebox{0.45}[0.4]{\includegraphics{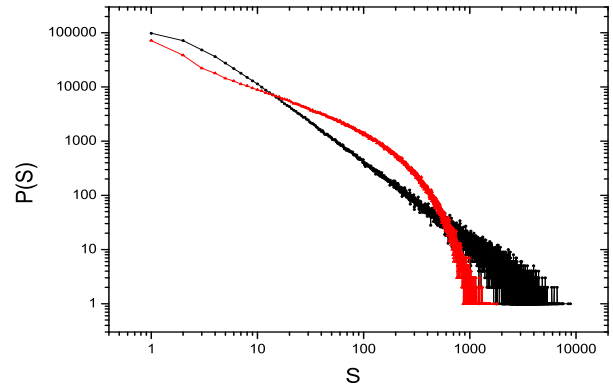}}
\caption{\label{fig:epsart} The distribution of avalanche size,
where $\varepsilon =0.01$. The red and black curves represent the
cases of Euclidean lattice and SF network respectively, where
$P(S)$ denotes the number of avalanches with give size $S$. The
maximal avalanche size in SF network is 8829 and the corresponding
quantity in Euclidean lattice is 1799. Both the two networks are
of 4900 vertices. The data shown here is obtained by $10^6+10^5$
iterations exclusive of initial $10^5$ time steps.}
\end{figure}

\begin{figure}
\scalebox{0.45}[0.4]{\includegraphics{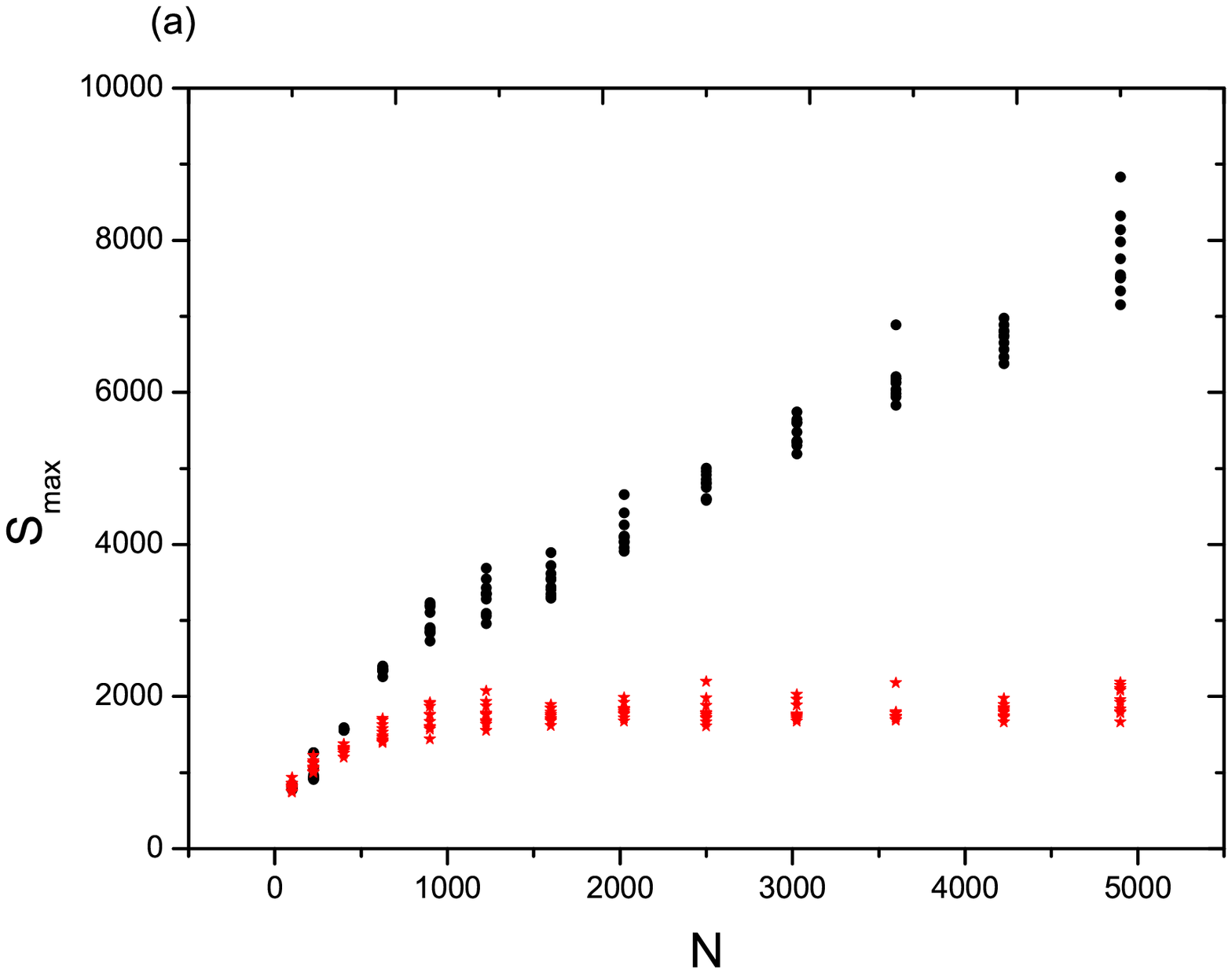}}
\scalebox{0.45}[0.4]{\includegraphics{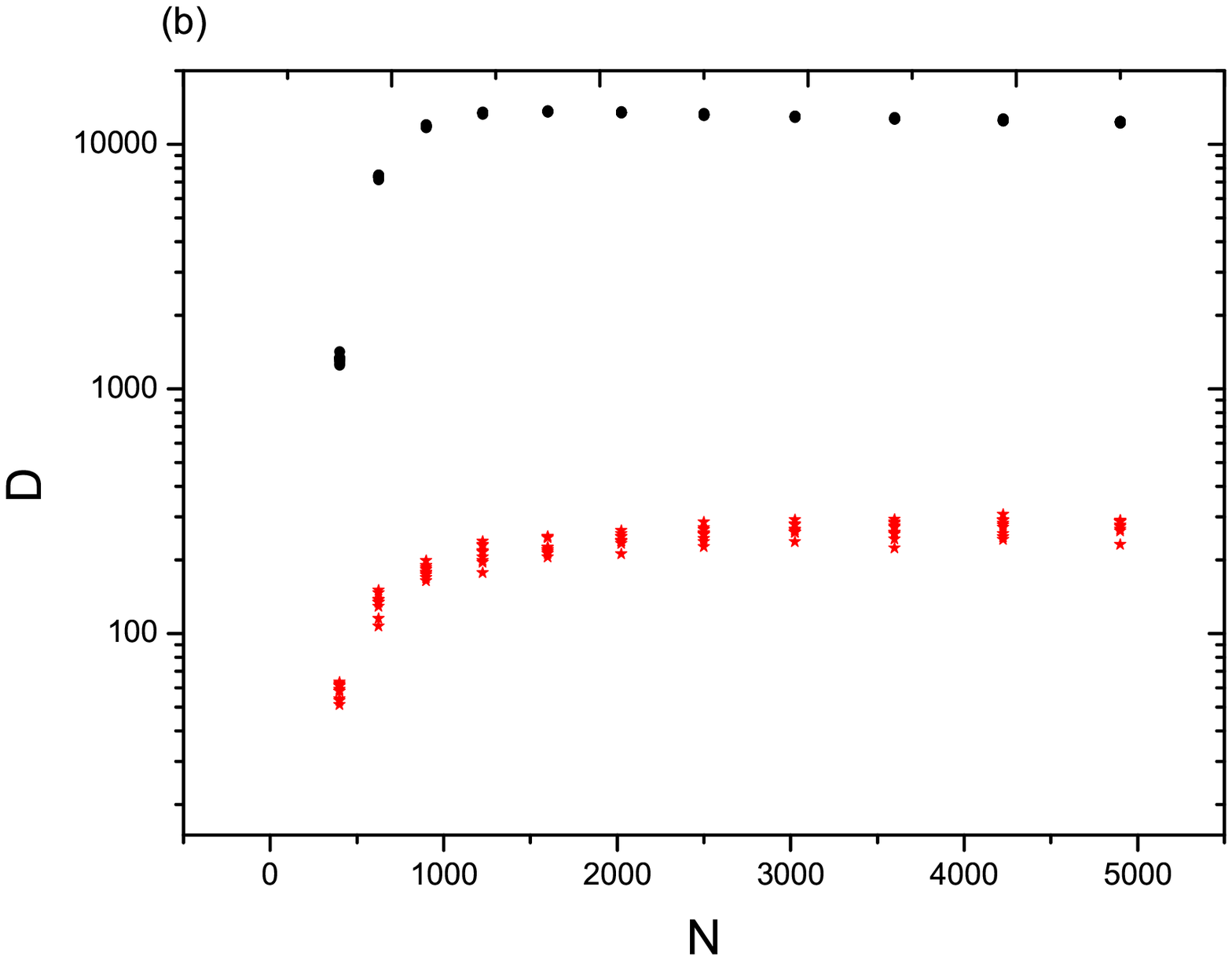}}
\caption{\label{fig:epsart} The degree and frequency of
catastrophes occurring in the two types of networks. In those two
figures, for a fixed $N$, the data are obtained by 10 independent
experiments, the results anent SF networks and Euclidean lattices
are represented by black (upper) dots and the red (lower) stars
respectively. {\bf a}, the size of greatest catastrophes in SF
networks is much more than that in Euclidean lattices, and the
disparity becomes greater and greater as the network-size
increases. {\bf b}, the frequency of catastrophes\cite{Zhou ex1}
occurring in SF networks is about 50 times higher than that in
Euclidean lattices. }
\end{figure}

\begin{figure}
\scalebox{0.45}[0.4]{\includegraphics{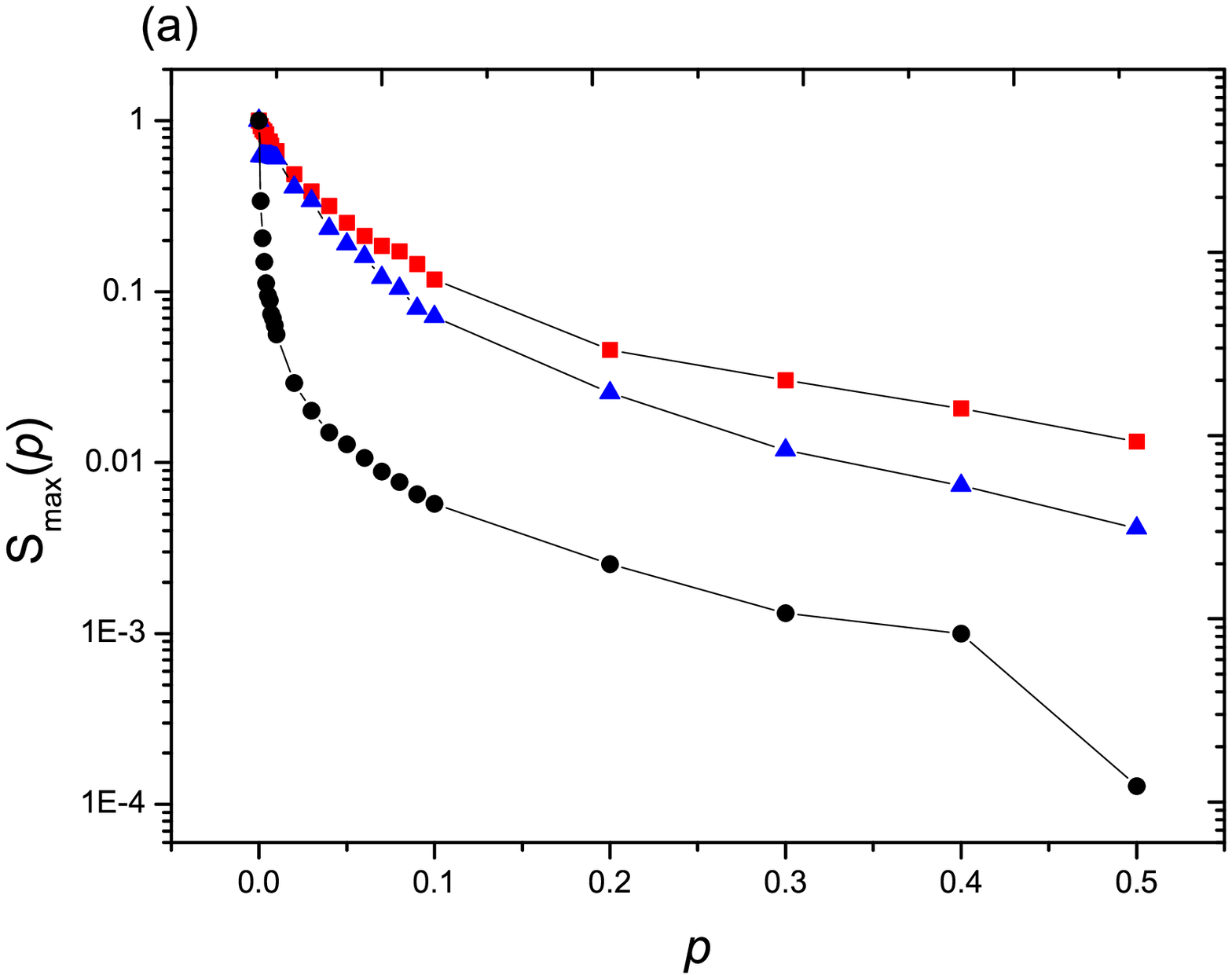}}
\scalebox{0.45}[0.4]{\includegraphics{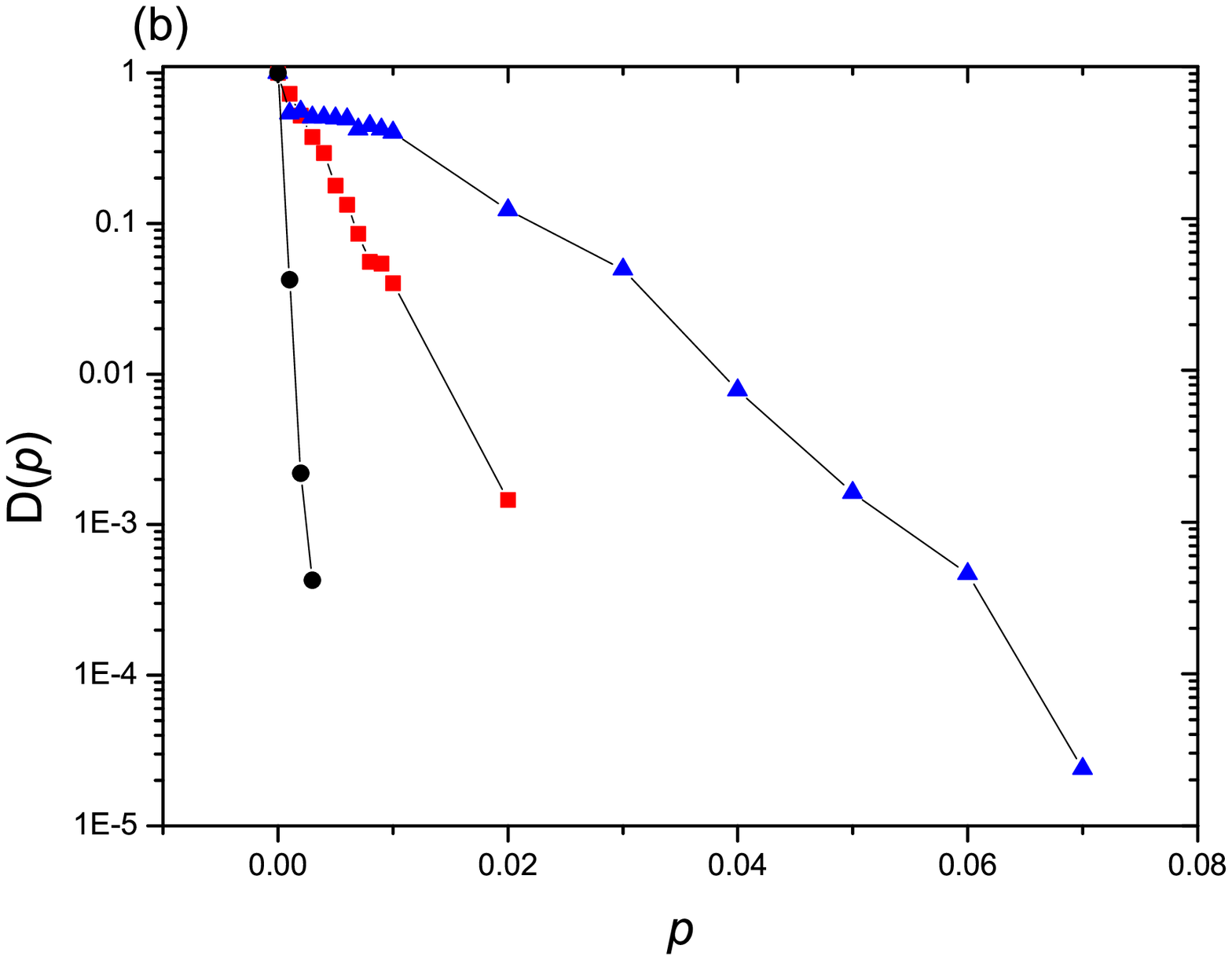}}
\scalebox{0.45}[0.4]{\includegraphics{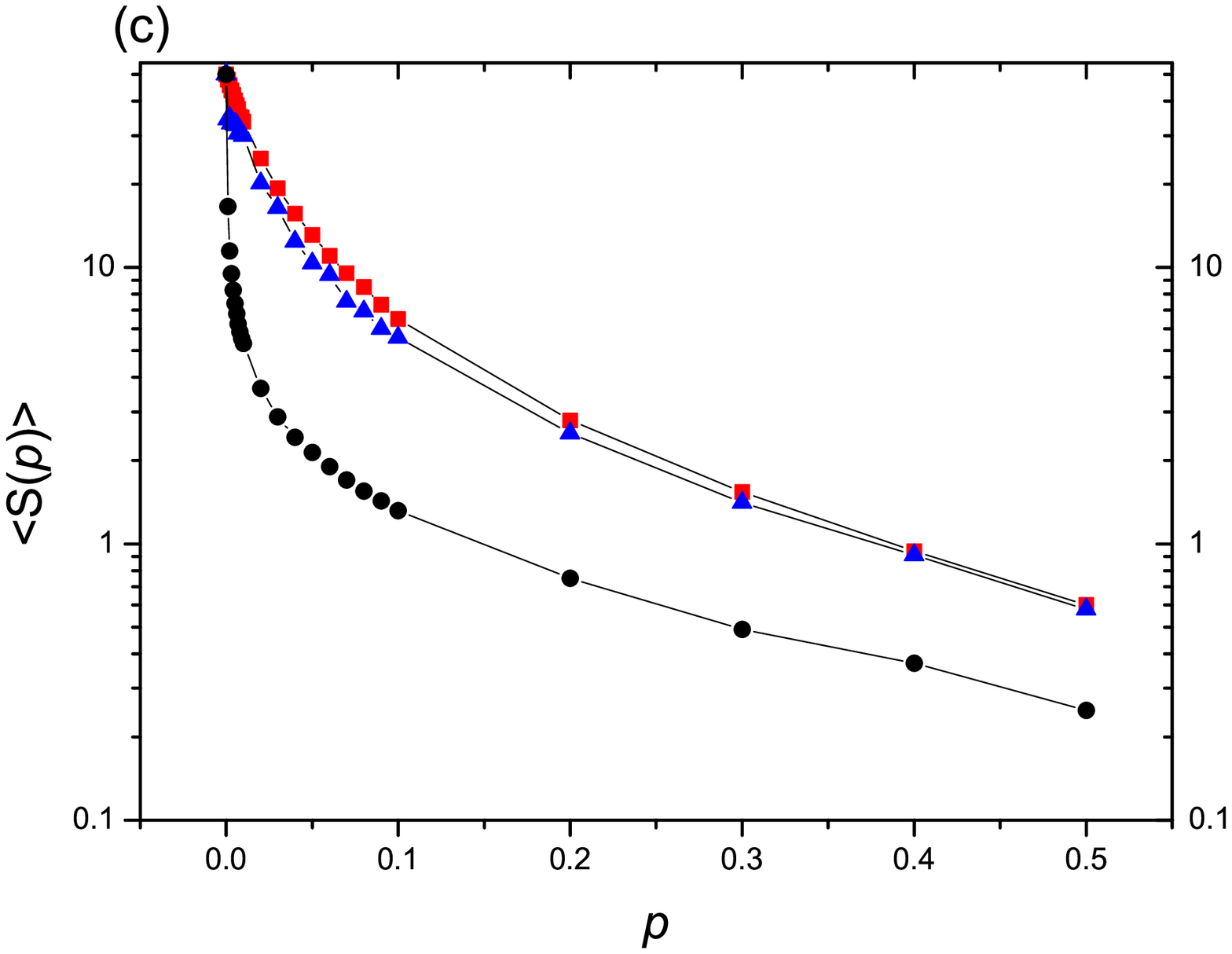}}
\caption{\label{fig:epsart} The catastrophes in networks under
vertex-protecting, where $N=4900$ and $\langle d\rangle$=4. For a
fixed $p$, the values shown here are the average over 10
independent experiments. The red squares$(\blacksquare)$, blue
triangles$(\blacktriangle)$ and black circles$(\bullet)$ denote
the performance of RSS (TSS) in Euclidean lattice, RSS in SF
network and TSS in SF network, respectively. Figure {\bf a}, {\bf
b} and {\bf c} present the maximal avalanche size $S_{max}$, the
number of catastrophes $D$ ($C=1000$) and the average avalanche
size $\langle S\rangle$ as functions of protecting rate $p$,
respectively. The values of $S_{max}$ and $D$ are normalized
(divided by $S_{max}(0)$ and $D(0)$). }
\end{figure}

In some networks like the internet, the vertex with greater degree
may have more throughput, thus {\it Goh}'s designation for
threshold value of vertices is reasonable\cite{God 0304}. But in
other networks (such as neural networks, social networks and so
on), there are not any evidences that the individual having more
neighbors is of greater endurance, thus it is worthy to study the
case that heterogeneous vertices are of the same threshold value.
Therefore, in our model, the threshold values are assigned to be
the same as $Z_x=Z=1$ for all vertices $x$, which is different
from {\it Goh}'s, It is notable that our designation for threshold
value is helpful to clearly understand how the topology of
networks affects the degree of catastrophe.

We are interested in the avalanche size $S$, which can be used to
measure the degree of catastrophe\cite{Zhou ex1} and defined as
the number of toppling events in a give avalanche (We set $S=0$ if
there is no toppling events occurring). Figure 1 shows a typical
result about the distribution of avalanche size. The Euclidean
lattice mentioned in this paper is a two-dimensional square
lattice with open boundary conditions, and the SF networks here
are gained by {\it Barab\'{a}si} and {\it Albert}'s method\cite{BA
99} with parameters $m_0=m=2$, thus both the two types of networks
are of average degree $\langle d\rangle \simeq 4$. One can see
that the distribution of avalanche size in SF network follows a
straight line for more than 3 decades, which indicates that there
is SOC in avalanche behavior. But the distribution of avalanche
size in Euclidean lattice, which is a power-law curve in the left
part followed by an approximately exponential truncation, is not
similar to that in SF network. Therefore, the dynamic behaviors in
those two types of networks are different. The presence of SOC in
the nonconservative OFC model has been controversial since the
very introduction of the model\cite{Debate 1} and it is still
debated\cite{Debate 2}. Since the main goal of this letter is to
study the catastrophes occurring in networks, we won't give
detailed experiment results and analysis on how the network
structure affects the existence of SOC, which will be given
elsewhere\cite{Zhou un}.

Getting to business, one can find that although the two networks
are of the same network-size\cite{Zhou ex2} and have the same
average avalanche size (apparently, $\langle S\rangle
=\frac{\langle \delta \rangle}{\varepsilon Z}=50$), the maximal
avalanche size $S_{max}$ in SF networks is much greater than that
in Euclidean lattice, which means the greatest catastrophe in SF
network is much more serious than that in Euclidean lattice. In SF
networks, the number of avalanches with its size larger than $C$
is 12291, and the corresponding quantity in Euclidean lattice is
231, which shows that the catastrophes occur much more frequently
in SF network than in Euclidean lattice. For the sake of reducing
the error, more experiments have been achieved, figure 2{\bf
a}\&2{\bf b} show the dependence of $S_{max}$ and the number of
catastrophes with the number of vertices $N$, which convictively
confirm the conclusions mentioned above.

Since there aren't any effective methods to put an end to
perturbations, it is worthwhile to study how to reduce the
catastrophes' degree. Here, for theoretic simplification, a
safeguard-strategy is defined as a vertices set $V_P$ with its
elements being protected and will not topple (mathematically
speaking, to protect a vertex $x$ here means to set its threshold
value $Z_x$ infinite). In this letter, two safeguard-strategies
are discussed, one is called random safeguard-strategy (RSS), and
the other one is called targeted safeguard-strategy (TSS). In the
former case, the vertices belonging to $V_P$ are randomly chosen;
in the latter one, the vertices with greater degree are chosen
preferentially. Since almost all the vertices in Euclidean lattice
are of the same degree, RSS and TSS in Euclidean lattice are not
discriminating.

We can always reduce the catastrophes' degree by protecting more
vertices, but this could bring in economical and technical
pressures. In order to roughly measure the economical and
technical expenses, a parameter $p$ called protecting rate is
defined as the proportion between the number of vertices being
protected and the total number of vertices: $p=|V_P|/N$. In figure
3, we report the experimental results about the different
safeguard-strategies. In figure 3{\bf a}, one can find that the
RSS in SF networks is a little more effective than that in
Euclidean lattices, and the TSS in SF networks is much more
effective than RSS. For example, if we want to reduce the maximal
avalanche size on SF network to a tenth by using TSS, then to
protect 0.5\% vertices is enough, but if we use RSS, the
protecting rate must be larger than 8\%. According to the results
shown in figure 3{\bf b}, if one want to eliminate the
catastrophes, at least 0.3\%, 2\% and 7\% vertices should be
protected by using TSS in SF network, RSS in Euclidean lattice and
RSS in SF network, respectively. One can also make out of figure
3{\bf c} that TSS in SF network is much more effective than RSS,
for example, if we set $p=0.01$, then the average avalanche size
will reduce to about 5 by using TSS, but the corresponding value
is about 30 for RSS. Altogether, the experimental results above
indicate that TSS is much more effective than RSS for SF networks.
The heterogeneity of vertices in SF networks is considered to be a
possible reason for these results.

In summary, we have found that the avalanche dynamic behaviors are
very sensitive to the topological structure of networks and the
catastrophes are much more serious in SF networks than in
Euclidean lattices. We have studied how to reduce the
catastrophes' degree, and have schemed out an effective strategy
(TSS) for SF networks, which may be of great significance in
practice. However, there are so many unanswered questions that
puzzle us. At the end of this letter, we will list part of those.

How does the topological structure of networks affect the
existence of SOC?

Is the highly skewed degree distribution the key reason why the
catastrophes in SF networks are much more serious than those in
Euclidean lattices?

How about the performance of limited safeguard-strategy\cite{Zhou
ex3}?

Are there other safeguard-strategies more effective than TSS for
SF networks?

This work has been partially supported by the State Key
Development Programme of Basic Research (973 Project) of China,
the National Natural Science Foundation of China under Grant
No.70271070 and the Specialized Research Fund for the Doctoral
Program of Higher Education (SRFDP No.20020358009)


\begin{thebibliography}{Review 1}
\bibitem{Zhou ex1} A catastrophe here is defined as the avalanche
with its size larger than an experiential lower bound $C$.
Although in this letter we suppose $C=1000=20\langle S\rangle$ ,
one can easily check that the phenomena are almost the same when a
proper value of $C$ is given ($C$ should be big enough but not too
big, the value between $10\langle S\rangle$ and $40\langle
S\rangle$ is preferable), thus our discussion do not depend upon
the value of $C$.
\bibitem{Zhou ex2} Networks are of the same network-size means
they have the same number of vertices and the same number of
edges.
\bibitem{Zhou ex3} Limited safeguard-strategy means for any vertex $x\in V_P$, the threshold value $Z_x$
is not infinite.
\end{thebibliography}
\end{document}